\documentclass[reprint,aip]{revtex4-1}
\pdfoutput=1
\usepackage{amsmath}
\usepackage{amssymb}
\usepackage{amsfonts}
\usepackage{bm}
\usepackage{xcolor}
\usepackage{graphicx} 
\usepackage{hyperref}
\hypersetup{
    bookmarks=true,         
    unicode=false,          
    pdftoolbar=true,        
    pdfmenubar=true,        
    pdffitwindow=false,     
    pdfstartview={FitH},    
    pdftitle={multi-scale model},
    pdfauthor={Eero Hirvijoki},
    pdfsubject={Kinetic theory},
    pdfcreator={Eero Hirvijoki},
    pdfproducer={GNU Make},
    pdfkeywords={Collisions} {multi-scale models},
    pdfnewwindow=true,
    colorlinks=true,
    linkcolor=blue,
    citecolor=blue,
    filecolor=blue,
    urlcolor=blue,  
}
\makeatother

\begin{document}

\title{A Langevin approach to multi-scale modeling}

\author{Eero Hirvijoki}
\affiliation{Princeton Plasma Physics Laboratory, Princeton, New Jersey 08543, USA}
\email{ehirvijo@pppl.gov}

\date{\today}

\begin{abstract}
In plasmas, distribution functions often demonstrate long anisotropic tails or otherwise significant deviations from local Maxwellians. The tails, especially if they are pulled out from the bulk, pose a serious challenge for numerical simulations as resolving both the bulk and the tail on the same mesh is often challenging. A multi-scale approach, providing evolution equations for the bulk and the tail individually, could offer a resolution in the sense that both populations could be treated on separate meshes, or different reduction techniques applied to the bulk and the tail population. In this letter, we propose a multi-scale method which allows us to split a distribution function into a bulk and a tail so that both populations remain genuine, non-negative distribution functions and may carry density, momentum, and energy. The proposed method is based on the observation that the motion of an individual test particle in a plasma obeys a stochastic differential equation, also referred to as a Langevin equation. This allows us to define transition probabilities between the bulk and the tail and to provide evolution equations for both populations separately. 
\end{abstract}

%\pacs{put Pacs here}%
%\keywords{put Keywords here}

\maketitle
 
%\section{Introduction}
\textit{Introduction} -- 
In many plasma physics problems of interest it is rather challenging to treat the full distribution functions. To make progress and build intuition, it is thus a common theoretical practice to seek to represent the bulk of the distribution functions as Maxwellians parametrized by the associated fluid quantities, and to advance the remainders of the distribution functions from kinetic principles. Alternatively, one could be interested in modeling the bulk and the tail populations on different meshes for purely computational reasons. For example, advanced algorithms that target energy and momentum preservation sometimes rely on structured, non-adaptive grids, and resolving both the bulk and the tail under such restrictions could quickly become computationally inefficient. Before proceeding to either direction, it could be beneficial to first have linearly independent evolution equations for the bulk and the tail distribution functions, on top of which reduced models or advanced numerical schemes could then be constructed.  

How to obtain such evolution equations in general remains largely unsolved. The moment approach, parametrizing the bulk with respect to the moments of the total distribution function and solving for the deviation kinetically, is limited in the sense that the deviation from the bulk is not allowed to carry any of the quantities used for parametrizing the bulk, which typically are density, momentum, and kinetic energy\cite{grad1949}. The popular $\delta f$-approach, on the other hand, becomes nontrivial when non-advective processes such as collisions are introduced\cite{Brunner_deltaf_1999PhPl}. More importantly, though, if the bulk distribution is evolved in time with moment equations, the same conditions are required from $\delta f$ as in the moment approach. Both methods thus have trouble handling long, anisotropic tails, and in neither can the deviating population be interpreted as a genuine, non-negative distribution function. Finally, simply splitting the distribution functions into a bulk and a tail population leads to linearly dependent equations that cannot be solved as such, unless either additional constraints, such as the moment approach conditions mentioned above, are imposed, or a closure is introduced to decouple the equations. 

In this letter, we propose a new multi-scale model to offer an alternative for modeling interactions between a bulk and a tail. We base our model on the observation that the motion of an individual test particle in a plasma obeys a stochastic differential equation, also referred to as a Langevin equation. This allows us to exploit a systematic probabilistic closure, and to derive linearly independent, solvable equations for the bulk and the tail populations that couple via transition probabilities. The resulting system of equations automatically respects the density, momentum, and energy conservation laws of the original system, and both the bulk and the tail are guaranteed to remain genuine, non-negative distribution functions. The generic recipe for implementing the model is discussed in detail and a simple numerical example for computing the transition probabilities is provided for demonstration purposes. 

In future, the new formulation could serve as a starting point for constructing reduced models where the bulk and the tail population may be treated with different methods, or allow for more efficient simulations of long anisotropic tails that interact with the bulk. Applications of the new model are expected, for example, in modeling runaway electrons in disrupting tokamak plasmas\cite{hirvijoki_fluid_kinetic:2018arXiv} or in estimating precipitation from reconnection-driven particle acceleration in magnetosphere studies. 

%\section{A probabilistic closure}
\textit{A probabilistic closure} -- 
In what follows, we shall use the full particle dynamics to provide an explicit example. Nothing, however, restricts applying the following idea to an already reduced full-f kinetic system which can be written down in the form of 
\begin{align}\label{eq:full-kinetic}
\frac{d f_{\alpha}}{dt}=\sum_{\beta}C_{\alpha\beta}[f_{\alpha},f_{\beta}],
\end{align}
with $d/dt$ a linear operator, $C$ a bilinear operator, and the subscripts $\alpha$ and $\beta$ referring to different species. To begin, we exploit the linearity of the operator $d/dt$, the bilinearity of the operator $C$, and split the distribution functions according to $f_{\alpha}=f_{\alpha 0}+f_{\alpha 1}$ for each species. Then we introduce the following formal split of \eqref{eq:full-kinetic}
\begin{align}
\frac{d f_{\alpha 0}}{dt}&=\sum_{\beta}C_{\alpha\beta}[f_{\alpha 0},f_{\beta 0}]+\sum_{\beta}C_{\alpha\beta}[f_{\alpha 0},f_{\beta 1}]-I_{\alpha},\\
\frac{d f_{\alpha 1}}{dt}&=\sum_{\beta}C_{\alpha\beta}[f_{\alpha 1},f_{\beta 1}]+\sum_{\beta}C_{\alpha\beta}[f_{\alpha 1},f_{\beta 0}]+I_{\alpha}.
\end{align}
It is straightforward to verify that the sum of the above two equations exactly reproduces \eqref{eq:full-kinetic} and, therefore, also any existing conservation laws. One should keep in mind, though, that this rudimentary splitting does not imply nor rely on existence of different time or space scales in the system. It merely provides an approach to split distribution functions while retaining linearly independent equations. In this respect, our multi-scale problem may be considered reduced to defining the as-yet-unknown interaction term~$I_{\alpha}$. 

Separately for each species, we propose the following expression
\begin{multline}
I(\bm{z},t)=\frac{f_{0}}{\tau}\left(1-\mathbb{E}[\mathbf{1}_{\Omega_0}(\bm{Z}_{t+\tau})|\bm{Z}_t=\bm{z}]\right)\\-\frac{f_{1}}{\tau}\mathbb{E}[\mathbf{1}_{\Omega_0}(\bm{Z}_{t+\tau})|\bm{Z}_t=\bm{z}],
\end{multline}
where $\tau$ is a characteristic, application specific time scale, the domain $\Omega_0$ refers to a ``bulk domain'', $\bm{z}$ denotes the phase-space location corresponding to $(\bm{x},\bm{v})$ in case of full particle dynamics, and $\bm{Z}_s$, with $s\in[t,t+\tau]$, denotes the trajectory of an individual particle in the phase-space. The operators $\mathbb{E}$ and $\mathbf{1}$ refer to an expectation value and an indicator function respectively. More specifically, $\mathbb{E}[\mathbf{1}_{\Omega_0}(\bm{Z}_{t+\tau})|\bm{Z}_t=\bm{z}]$ is the probability for finding a particle with an initial position $\bm{z}$ at time $t$ within the domain $\Omega_0$ after the time interval $\tau$. Thus, as appearing in the equation for $f_1$, the interaction term adds particles from the bulk at a rate that is proportional to the number of bulk particles available and the probability of a bulk particle leaving the bulk domain after time $\tau$. At the same time, it depletes particles from $f_1$ at a rate that is proportional to number of particles available in the $f_1$ population and the probability of such particle to remain within the bulk domain. As appearing in the equation for $f_0$, the action is opposite. Note that the phase-space location of the particles does not change.

The chosen form of the interaction term guarantees the non-negativity of $f_{0}$ and $f_{1}$, given that the linear advection operator $d/dt$ and the bilinear operator $C$ do so. For any point $\bm{z}_{\star}$ where $f_0(\bm{z}_{\star},t)=0$, the contribution from the interaction term to the evolution of $f_0$ is $f_1(\bm{z}_{\star},t)\mathbb{E}[\mathbf{1}_{\Omega_0}(\bm{Z}_{t+\tau})|\bm{Z}_t=\bm{z}_{\star}]/\tau\geq 0$, increasing the value of $f_0$. Similarly, if $f_1(\bm{z}_{\star},t)=0$, the contribution from the interaction term to the evolution of $f_1$ is $f_0(\bm{z}_{\star},t)(1-\mathbb{E}[\mathbf{1}_{\Omega_0}(\bm{Z}_{t+\tau})|\bm{Z}_t=\bm{z}_{\star}])/\tau\geq 0$, increasing the value of $f_1$. This stems from the fact that $\mathbb{E}[\mathbf{1}_{\Omega_0}(\bm{Z}_{t+\tau})|\bm{Z}_t=\bm{z}]\in[0,1]$. Both $f_0$ and $f_1$ can thus be interpreted as genuine distribution functions. We also expect the proposed interaction term to provide a stable splitting scheme void of unphysical oscillations and exponentially growing modes: assuming $f_0$ and $f_1$ to be driven only with the interaction term with a fixed value for the expectation, both $f_0$ and $f_1$ would relax exponentially with a time-scale $\tau$ to an equilibrium determined by the initial values for $f_0$ and $f_1$ and the expectation value and its complement. 

To determine the evolution of $\bm{Z}_s$ during $s\in [t,t+\tau]$, we turn to physics. It is well known that the motion of an individual test-particle in a given background plasma obeys a stochastic differential equation, or a so-called Langevin equation\cite{karatzas2012}. This is typically true, regardless of the level of approximation that is used for estimating the test-particle trajectories. For the sake of finding a method for evaluating $\mathbb{E}[\mathbf{1}_{\Omega_0}(\bm{Z}_{t+\tau})|\bm{Z}_t=\bm{z}]$ efficiently, we will also make the assumption that the quantities needed in determining the test-particle motion can be considered constant in time during $s\in [t,t+\tau]$, regardless of the model used for estimating the trajectory. The motivation behind this requirement will be clarified soon. Hence the effects of phenomena that occur in time scales faster that $\tau$ would need to be accounted for via, e.g., the quasilinear approximation, to restrict only the slowly varying dynamics to contribute to the deterministic test-particle motion. Alternatively, one could choose the time-scale $\tau$ to match the characteristic time scale of interest. Note that the above assumptions do not impose conditions on the scales present in $f_0$ or $f_1$, only on how the transition probability between the two populations is estimated. 

Considering the full particle dynamics as an example, the stochastic motion in It\^o convention would then follow
\begin{align}
d\bm{X}_s&=\bm{V}_sds,\\
d\bm{V}_s&=\frac{e}{m}[\bm{E}(\bm{X}_s,t)+\bm{V}_s\times\bm{B}(\bm{X}_s,t)]ds\nonumber\\
&\qquad+\bm{\mu}(\bm{Z}_s,t)ds+\bm{\sigma}(\bm{Z}_s,t)\cdot d\bm{W}_s,
\end{align}
where the electromagnetic fields, $\bm{E}$ and $\bm{B}$, and the coefficients $\bm{\mu}$ and $\bm{\sigma}$ are now estimated constant on the time-scale $\tau$ and evaluated at time $t$ but at the particle position $\bm{Z}_s$, and $\bm{W}_s$ is a standard vector-valued Wiener process (see Ref\cite{vanKampen_1981JSP} for a discussion on It\^o and Stratonovich conventions). For test-particle motion, the $\bm{\mu}$ and $\bm{\sigma}$ coefficients should include at least the test-particle Coulomb collision operator 
\begin{align}
C_{\alpha\beta}[f_{\alpha},f_{\beta 0}]=\frac{\partial}{\partial\bm{v}}\cdot\left(\mathbb{D}_{\alpha\beta 0}\cdot\frac{\partial f_{\alpha}}{\partial\bm{v}}-\bm{F}_{\alpha\beta 0}f_{\alpha}\right),
\end{align}
providing the relations
\begin{align}
\bm{\mu}_{\alpha}&=\sum_{\beta}\left(\bm{F}_{\alpha\beta 0}+\frac{\partial}{\partial\bm{v}}\cdot\mathbb{D}_{\alpha\beta 0}\right),\\ 
\bm{\sigma}_{\alpha}\cdot\bm{\sigma}_{\alpha}^{\text{T}}&=2\sum_{\beta}\mathbb{D}_{\alpha\beta 0},
\end{align}
with superscript $\text{T}$ referring to a transpose of a matrix. Additionally, if fast electromagnetic waves were to be expected, quasilinear diffusion, accounting for wave-particle interaction, could be included as well.

Given dynamics for a test-particle, the expectation value can now be expressed as
\begin{align}
\mathbb{E}[\mathbf{1}_{\Omega_0}(\bm{Z}_{t+\tau})|\bm{Z}_t=\bm{z}]=\int \mathbf{1}_{\Omega_0}\left(\bm{Z}_{t+\tau}\right)d\mu\left(\bm{W}_{\tau}\right),
\end{align}
where $\mu(\bm{W}_{\tau})$ refers to the formal Borel measure the standard Wiener process $\bm{W}_s$ generates during the time interval $\tau$. Given an arbitrary $\tau$, the above expectation value could be computed with forward Monte Carlo simulation of multiple stochastic trajectories $\bm{Z}_s$ and taking the statistical average with respect to the end condition $\bm{Z}_{t+\tau}\in\Omega_0$. It is also clear that, since only single-particle characteristics are required, reduced models could be introduced whenever applicable. For example, the guiding-center dynamics could be used instead of the full particle motion to speed up the evaluation of approximate particle characteristics. Most importantly, although the forward Monte Carlo simulation of test-particle characteristics can be trivially parallelized, it can be avoided entirely. Given a time independent stochastic differential equation in its It\^o form, efficient methods to compute expectation values deterministically exist. One may use, for example, the adjoint formulation\cite{karney_current_1986,Liu_2016_adjoint} or the more formal link via the Feynman-Kac equation\cite{karatzas2012} as was demonstrated in Refs\cite{Zhang-DiegoPoP2017-backwards-monte-carlo,bormetti_callegaro_livieri_pallavicini_2018}. The approach exploiting the Feynman-Kac formula, and other practical matters, will be explained next. 

%\section{Practical simplifications}
\textit{Practical considerations} -- 
Our assumption to treat the background plasma independent of time during the interval $s\in[t,t+\tau]$ leads to a useful observation. If we define a quantity
\begin{align}
\Phi(\bm{z};s)=\mathbb{E}[\mathbf{1}_{\Omega_0}(\bm{Z}_{t+\tau})|\bm{Z}_s=\bm{z}],
\end{align}
we first of all find that $\Phi(\bm{z};t)=\mathbb{E}[\mathbf{1}_{\Omega_0}(\bm{Z}_{t+\tau})|\bm{Z}_t=\bm{z}]$ corresponds to the expectation we need for computing the interaction term, and that $\Phi(\bm{z};t+\tau)=\mathbf{1}_{\Omega_0}(\bm{z})$. The latter follows from the test particle not moving anywhere in zero time. These two observations allow us to write
\begin{align}
\Phi(\bm{z};t)=\mathbb{E}[\Phi(\bm{Z}_{t+\tau};t+\tau)|\bm{Z}_t=\bm{z}],
\end{align}
which is a special case of the Feynman-Kac formula. Moreover, due to the Markovian property of the stochastic differential equation, we may create a partition for the interval $\tau$ according to $\{t=s_0,s_1,...,s_N=t+\tau\}$, and compute $\Phi(\bm{z};t)$ backwards in the time starting from $\Phi(\bm{z};t+\tau)=\mathbf{1}_{\Omega_0}(\bm{z})$ according to the rule
\begin{align}
\Phi(\bm{z};s_{n-1})=\mathbb{E}[\Phi(\bm{Z}_{s_n};s_n)|\bm{Z}_{s_{n-1}}=\bm{z}].
\end{align}

To efficiently evaluate the terms $\Phi(\bm{z};s_{n})$, the partition $\{s_n\}_{n=0}^N$ should be created with equal subintervals $\Delta s$ and the sub-interval chosen so that the variance in the particle velocity due to stochastic scattering during $\Delta s$ would not be too large. This way the particle's phase-space position after the interval $\Delta s$ can be estimated to a good accuracy by first integrating along the deterministic trajectory and then providing a ``kick'' in the velocity with respect to the change in the Wiener process. The possibly costly forward Monte Carlo simulation is now replaced by integrating along the deterministic orbit and then taking the statistical average with respect to the distribution of the kicks. The advantage of dividing $\tau$ into equal subintervals $\Delta s$ is that, since the background does not change in time, in each iteration of the functions $\Phi(\bm{z};s_{n})$ the same deterministic orbit can be used for a given $\bm{z}$. This way, the typically costly interpolation of the electromagnetic fields needed in forward Monte Carlo approach is avoided.

Using the full particle dynamics as an example, the previous discussion results in the following recipe: given a position $\bm{z}$, the particle position $\bm{Z}_{\Delta s}=(\bm{X}_{\Delta s},\bm{V}_{\Delta s})$ after the interval $\Delta s$ is estimated, e.g.,  with the Euler-Maruyama discretization of the stochastic differential equations according to
\begin{align}
\bm{X}_{\Delta s}&=\bm{x}+\int_{0}^{\Delta s}\frac{d \bm{x}}{ds}ds\\
\widetilde{\bm{V}}_{\Delta s}&=\bm{v}+\int_{0}^{\Delta s}\frac{d \bm{v}}{ds}ds,\\
\bm{V}_{\Delta s}&=\widetilde{\bm{V}}_{\Delta s}+\bm{\sigma}\left(\bm{X}_{\Delta s},\widetilde{\bm{V}}_{\Delta s},t\right)\cdot\bm{\xi}\sqrt{\Delta s},
\end{align}
where $\bm{\xi}\sim\mathcal{N}(\bm{0},\mathbf{I})$ obeys the standard multivariate normal distribution and the deterministic trajectory obeys the ordinary differential equations
\begin{align}\label{eq:ode}
\frac{d\bm{x}}{ds}&=\bm{v},\\
\frac{d\bm{v}}{ds}&=\frac{e}{m}[\bm{E}(\bm{x},t)+\bm{v}\times\bm{B}(\bm{x},t)]+\bm{\mu}(\bm{x},\bm{v},t).
\end{align}
As the particle position $\bm{Z}_{\Delta s}(\bm{\xi})$ depends only on the change $\Delta\bm{W}=\bm{\xi}\sqrt{\Delta s}$ in the Wiener-process over the time interval $\Delta s$, and not on any of its intermediate realizations, the iteration rule becomes 
\begin{align}\label{eq:probability}
\Phi(\bm{z},s_{n-1})=\int_{\mathbb{R}^3}\Phi(\bm{Z}_{\Delta s}(\bm{\xi}),s_n)\frac{\exp[-\bm{\xi}^2/2]}{(2\pi)^{3/2}}d\bm{\xi}.
\end{align}
Alternative, more sophisticated stochastic discretization schemes could be used as well, such as the Milstein method, though higher order methods tend to quickly grow somewhat complex, typically requiring derivatives of the coefficient $\sigma$ as well as sampling of the so-called area integrals of type $\int_0^{\Delta s}d\bm{W}_{s'}\int_0^{s'}d\bm{W}_{s''}$ to estimate $\bm{Z}_{\Delta s}(\bm{\xi})$. For a thorough introduction to higher order stochastic discretization methods, see, e.g., Refs \cite{Gardiner_handbook:1994,Gaines_Lyons_1994,Ryden_2001,Dimits_et_al_2013JCoPh}.

The iterative approach requires the evaluation of only one deterministic trajectory per location $\bm{z}$ but the averaging over the stochastic kicks at the end of the deterministic trajectory necessitates information of the function $\Phi$ in points $\bm{Z}_{\Delta s}(\bm{\xi})$. In practice, a mesh would be required for the iterative evaluation of the expectation value. From a computational point-of-view, it would be beneficial not to resort to such measures, and instead to be able to evaluate the expectation value based on only the information of that one deterministic particle trajectory. This can be achieved if the time-interval $\tau$ itself is chosen short enough. In this case, it is enough to take only one iteration, leading to the estimate
\begin{align}\label{eq:one-step}
\Phi(\bm{z};t)=\int_{\mathbb{R}^3}\mathbf{1}_{\Omega_0}(\bm{Z}_{\Delta s}(\bm{\xi}))\frac{\exp[-\bm{\xi}^2/2]}{(2\pi)^{3/2}}d\bm{\xi},
\end{align}
with the time step given by $\Delta s =\tau$. Since the indicator function is given, no mesh is required for evaluating $\Phi$, only the definition of the domain $\Omega_0$. In purely numerical applications, one would choose $\Omega_0$ to correspond to the meshed domain for the population $f_0$. In applications that seek fluid modeling for $f_0$ and kinetic modeling for $f_1$, a convenient definition would be a sphere in velocity space, shifted from the origin by the mean-flow velocity of the bulk population, with a radius given in units of the local thermal velocity. 

Further simplifications to the proposed multi-scale method are easy to adopt. Since the goal is to estimate whether the particle ends its trajectory within the bulk domain, one possibility would be to assume the bulk quantities needed for evaluating the coefficients $\bm{\mu}$ and $\bm{\sigma}$ constant along the particle trajectory, so that the actual spatial displacement of the particle could be ignored entirely. This approach could be applied for example in the Tokamak core, where the background temperature and density are approximately constant on magnetic flux-surfaces and the particle approximately follows the magnetic field lines. If further computational efficiency is necessary, one could estimate also only the particles trajectory in energy in which case the statistical averaging over the kicks in the particle velocity would reduce into a one-dimensional integral. 

%\section{An example}
\textit{An example} -- 
To provide a demonstration as simple as possible, we consider a homogenous plasma free of electric and magnetic fields so that only the collisional motion of a test particle in energy is relevant. This is a convenient one-dimensional model problem and, if extended to include pitch-angle dynamics and electric field, the model could be used to efficiently estimate, e.g., escape probabilities of runaway electrons in tokamak plasmas, as has been recently demonstrated\cite{Zhang-DiegoPoP2017-backwards-monte-carlo}. If extended further to include also one spatial dimension, it could become relevant for modeling, e.g., the fuel dilution effects of thermal alpha particles in inertial fusion targets, a problem that has been shown to benefit from a multi-mesh approach\cite{Peigney_et_al_2014JCoPh}.

\begin{figure}[!h]
\includegraphics[width=\columnwidth]{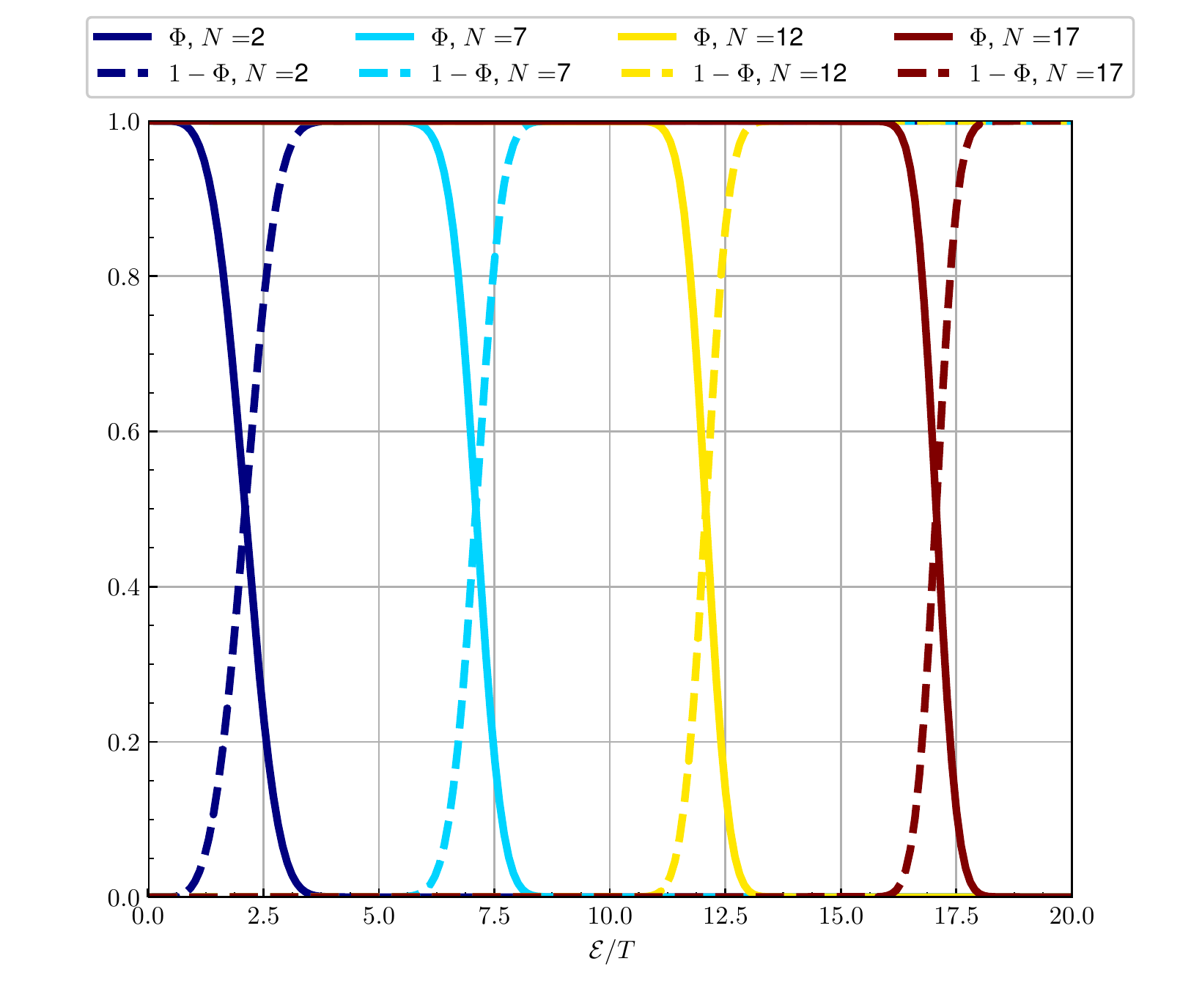}
\caption{\label{fig:sinksource} The transition probability and its complement as a function of energy normalized to the temperature for $M=10$ and different $N$. Notice the steepening of $\Phi$ with increasing $N$ which reflects that the particles become less collisional at higher energies.}
\end{figure}

Accounting for test-particle Coulomb collisions, the stochastic differential equation in energy $\mathcal{E}_{\alpha}$ for a particle of species $\alpha$ is given by
\begin{align}
d\mathcal{E}_{\alpha}=\mu_{\alpha}dt+\sigma_{\alpha}dW,
\end{align}
where the drift and the variance coefficients are 
\begin{align}
\mu_{\alpha}&=\sum_{\beta}3\nu_{\alpha\beta}T_{\beta}-2\nu_{\alpha\beta} \mathcal{E}_{\alpha}+2\frac{\partial \nu_{\alpha\beta}}{\partial \mathcal{E}}T_{\beta}\mathcal{E}_{\alpha},\\
\sigma^2_{\alpha}&=4\sum_{\beta}\nu_{\alpha\beta}T_{\beta}\mathcal{E}_{\alpha},
\end{align}
the energy scattering frequency $\nu_{\alpha\beta}$ is 
\begin{align}
\nu_{\alpha\beta}=\frac{\sqrt{m_{\beta}}n_{\beta}e_{\alpha}^2 e_{\beta}^2\ln\Lambda}{4\sqrt{2}\pi\varepsilon_0^2 m_{\alpha}T_{\beta}^{3/2}}\sqrt{\frac{m_{\alpha}}{m_{\beta}}\frac{T_{\beta}}{\mathcal{E}_{\alpha}}}\Psi\left(\sqrt{\frac{m_{\beta}}{m_{\alpha}}\frac{\mathcal{E}_{\alpha}}{T_{\beta}}}\right),
\end{align}
$\Psi(x)$ is the Chandrasekhar function
\begin{align}
\Psi(x)=\frac{\text{erf}(x)-2\pi^{-1/2}x\exp(-x^2)}{2x^2},
\end{align}
and temperature $T_{\beta}$ is in units of energy. We consider the bulk domain to be $\Omega_0=[0,N T_{\alpha}]$ with $N$ a positive integer. For the stochastic scattering not to be too large, the time-scale $\tau$ is chosen to correlate with the collision time scale according to
\begin{align}
M\tau \max\left(\sum_{\beta}\nu_{\alpha\beta}\right)=1,
\end{align}
with $M$ a positive integer. Considering a test electron colliding with an electron background with density $n_e=10^{20}\ [1/\text{m}^3]$ and temperature $T_e=5\ \text{keV}$, the transition probability and its complement are illustrated in Fig.\ref{fig:sinksource} for $M=10$ and different $N$ using formula \eqref{eq:one-step} in the one-dimensional case. Multiplying the corresponding curves with the $f_0$ and $f_1$ distributions would provide the terms needed to construct the interaction term for the problem. While the example is simple, it demonstrates two characteristic features of the method. (i) The evaluation of $\Phi$ is computationally appealing, in the test case requiring one evaluation of $\mu$ and $\sigma$ and two evaluations of the error function. (ii) The shape of the transition probability function depends on the definition of the boundary, steepening in the model problem for larger values of energy as the collisionality decreases at higher test-particle energies.

%\section{Summary}
\textit{Summary} --
We have presented a mathematically rigorous formalism to obtain linearly independent evolution equations for both a bulk and a tail population. In contrast to traditional approaches, the proposed method preserves the non-negativity of both the bulk and the tail population, predicting them as genuine distribution functions. We have also demonstrated the flexibility of the new method with respect to the degree of computational effort available for executing the algorithm. The superior efficiency of the critical part of the algorithm has been demonstrated also in the context of computing escape probabilities for runaway electrons\cite{Zhang-DiegoPoP2017-backwards-monte-carlo}. The new formalism could turn out especially valuable in providing consistent formulations for mixed reduced models, where the tail population is treated kinetically and the bulk as a fluid, or in numerical simulations where the distribution functions are expected to demonstrate long non-isotropic tails and two separate meshes are unavoidable for accurate treatment of both the bulk and the tail.

The Author would like to thank Luis Chac\'on, William Taitano, Diego del-Castillo-Negrete, Guannan Zhang, Chang Liu, Joshua Burby, and Dylan Brennan for useful discussions. Also the funding secured by Amitava Bhattacharjee is greatly appreciated. This work was supported by the U.S. Department of Energy Contract No. DE-AC02-09-CH11466 and grant {DE-SC0016268}. The views and opinions expressed herein do not necessarily reflect those of the U.S. Department of Energy.

\bibliographystyle{apsrev4-1}
\bibliography{langevin_bibfile}    
\end{document}